\documentclass{ifacconf}

\usepackage{graphicx}      
\usepackage{natbib}        
\usepackage{epstopdf}
\usepackage{amsmath}
\usepackage{amsfonts}
\usepackage{dsfont}
\usepackage{amssymb}
\usepackage{color}

\newtheorem{definition}{{Definition}} 
\newtheorem{remark}{{Remark}} 
\newtheorem{theorem}{Theorem}
\newtheorem{corollary}{Corollary}
\newtheorem{proposition}{Proposition}
\newtheorem{assumption}{Assumption}
\newtheorem{lemma}{Lemma}
\begin{document}
\begin{frontmatter}

\title{Virtual Differential Passivity based Control for Tracking of Flexible-joints Robots}


\author[First,Second]{Rodolfo Reyes-B\'aez}
\author[First,Second]{Arjan van der Schaft} 
\author[First,Third]{Bayu Jayawardhana} 

\address[First]{Jan C. Willems Center for Systems and Control,\\ University of Groningen, The Netherlands}

\address[Second]{Johann Bernoulli Institute for Mathematics and Computer Science, University of Groningen, P.O. Box 407, 9700 AK, Groningen, The Netherlands (\{r.reyes-baez, a.j.van.der.schaft\}@rug.nl)}

\address[Third]{Engineering and Technology Institute Groningen (ENTEG), University of Groningen,  Nijenborgh 4, 9747AG, The Netherlands (b.jayawardhana@rug.nl)}

\begin{abstract}                
	Based on recent advances in contraction methods in systems and control, in this paper we present the virtual differential passivity based control  (v-dPBC)	technique. This is a constructive  design method that combines the concept of virtual systems and of differential passivity. We apply the method to the  tracking control problem of flexible joints robots (FJRs) which are formulated in the port-Hamiltonian (pH) framework. Simulations on a single flexible joint link are presented for showing the performance of a controller obtained with this approach. 
\end{abstract}


\end{frontmatter}

\section{Introduction}
The problem of control of rigid robots has been widely studied  since they are  instrumental in modern manufacturing systems. However,  the elasticity phenomena  in the joints can not be neglected for accurate position tracking as reviewed in  \cite{tomei1995tracking}.  For every joint that is actuated by a motor, we need two degrees of freedom joints instead of one. Such FJRs are therefore \emph{underactuated}. In \cite{spongflexible} two state feedback control laws based on  feedback linearization and singular perturbation are presented for a simplified model. Similarly in \cite{canudas} a dynamic feedback controller for a more detailed model is presented. In \cite{loria1995tracking} a computed-torque controller for FJRs is designed, which does not need \emph{jerk} measurements.  In \cite{ortega-regulacion} and \cite{Brogliato40}  passivity-based control (PBC) schemes are proposed. The first one is an observer-based controller which requires only motor position measurements. In the latter one a PBC controller is designed and compared with backstepping and decoupling techniques. For further details on PBC of FJRs we refer to \cite{ortega2013passivity} and references therein. In \cite{astolfi2003immersion}, a global tracking controller based on the I\&I method is introduced. From a more practical point of view, in \cite{albu2007unified}, a torque feedback is embedded into the passivity-based control approach, leading to a full state feedback controller; with this acceleration and jerk measurements are not required. In a recent work of  \cite{Sofia}, they design a dynamic controller which solves the global position tracking problem  of FJRs based only on measurements of link and joint positions. All controllers mentioned above are for the second order Euler-Lagrange (EL) systems. Most of these schemes are based on the selection of a suitable storage function that together with the dissipativity of the closed-loop system,  ensures the convergence state trajectories to the desired solution.

As an alternative to the EL formalism, the pH framework has been introduced in \cite{vanderschaft1995}. The main characteristics of the pH framework are the existence of a Dirac structure (connects geometry with analysis),  port-based network modeling and  a \emph{clear physical energy interpretation}. For the latter part, the energy function can directly be used to show the dissipativity and stability property of the systems. 
Some  set-point  controllers have been proposed for FJRs modeled as pH systems. For instance in \cite{Borja2014flexible} the EL-controller for FJRs in \cite{ortega2013passivity} is adapted and interpreted in terms of  Control by Interconnection\footnote{We refer interested readers on  CbI to   \cite{castanos}.} (CbI). In \cite{chinoflexiblejoint}, they propose an Interconnection and Damping Assignment  PBC (IDA-PBC)\footnote{For IDA-PBC technique see also \cite{escobar}.} scheme, where the controller is designed with respect to the pH representation of the  EL-model  in  \cite{albu2007unified}.  For the tracking control problem of FJRs in the pH framework, to the best of our knowledge, the only result is the one in \cite{hilde}, where a singular perturbation approach is considered. The key result is that, both, the slow and fast dynamics are fully-actuated pH systems, so that we can apply directly rigid robots controllers.

In this work we extend our previous  results in  \cite{RodoAutomatica2017,Reyes-BaezIFAC2017}, on  v-dPBC of fully-actuated mechanical port-Hamiltonian systems, to solve the tracking problem of FJRs viewed as pH systems. This  method  relies on the \emph{contraction} properties of the so-called virtual systems,  \cite{forni,arjan2013differentialpassivity,pavlov2017convergent,slotinecontraction,sontag2010contractive,wang}.   Roughly speaking, the  method\footnote{The  use of virtual systems for control design  was already considered in \cite{jouffroy} and \cite{manchester2015unifying}.} consists in designing a control law for a  virtual system associated to a FJR, such that it is differentially passive in the closed-loop and has a desired steady-state behavior.  Then, the FJR in closed-loop with above controller tracks the virtual system's steady-state. 

The paper is organized as follows: In Section 2, the theoretical preliminaries on differential incremental methods, their relation with virtual systems and the v-dPBC methodology are presented. Section 3 deals with some structural properties of mechanical pH systems and the explicit pH model of FJRs, together with its associated virtual mechanical system. A trajectory tracking v-dPBC scheme for FJRs  is presented in section 4. In order to show the performance of a controller obtained with the proposed method, simulation results are presented. Finally, in Section 5 conclusions and future research are stated.

\section{Contraction, Differential passivity and virtual systems}\label{sII}
	
	In this paper, we adopt \emph{the differential Lyapunov framework} for contraction analysis as in the paper \cite{forni}, which unifies different approaches.  Some arguments will be omitted due to space limitation.
	
	Let $\Sigma$ be a nonlinear control system with state space $\mathcal{X}$ be the state-space of dimension $N$, affine in the input $u$, 
	\begin{equation}
	\Sigma:\left\{ \begin{array}{llc}
	\dot{{x}}={f}(x,t)+\sum_{i=1}^{n}{g}_i(x,t){u}_i,\\
	y_i=h_i(x,t), \quad i\in\{1,\cdots,n\},\\
	\end{array}
	\right.	
	\label{eq:controlsystem}
	\end{equation}
	where $x\in\mathcal{X}$, ${u}\in\mathcal{U}\subset\mathds{R}^n$ and $y\in\mathcal{Y}$. The vector fields ${f},{g}_i:\mathcal{X}\times\mathds{R}_{\geq 0}\rightarrow T\mathcal{X}$ are  assumed to be smooth and $h_i:\mathcal{X}\rightarrow \mathds{R}$, for $i\in\{1,\cdots,n\}$. The input space $\mathcal{U}$ and output space $\mathcal{Y}$ are assumed to be open subsets of $\mathds{R}^n$.

	Given two initial states $x(t_{1})=x_{10}$ and $x(t_2)=x_{20}$, take any  forward invariant coordinate neighborhood $\mathcal{C}$ of $\mathcal{X}$, containing  $x(t_0)=x_{10}$ and $x_{20}$. Consider a regular smooth curve  $\gamma:I\rightarrow \mathcal{C}$, $I:=[0,1]$, such that  $\gamma(0)=x_{10}$ and  $\gamma(1)=x_{20}$.  Let $t\in[t_0,T]\mapsto x(t)=\psi_{t_0}^{u}(t,\gamma(s))$ be the solution to \eqref{eq:controlsystem} from the initial condition $\gamma(s)$, at time $t_0$, corresponding to the family of input functions $t\in[t_0,T]\mapsto u(t)=\varrho_{t_0}(t,s)$, for $s\in I$. The differential in direction $\frac{\partial}{\partial t}$ at a fixed $s$ represents the time derivative. 	
	The time derivative of  $\psi_{t_0}^{u}:\mathcal{X}\times\mathds{R}_{\geq 0}\rightarrow T_{\gamma(s)} \mathcal{X}$  satisfies, 
	\begin{equation}
	\begin{split}
	\frac{\partial \psi_{t_0}^{u}}{\partial t}(t,\gamma(s))&=f(\psi_{t_0}^{u}(t,\gamma(s)),t)\\&+\sum_{i=1}^{n}g_i(\psi_{t_0}^{u}(t,\gamma(s)),t)\varrho_{i,t_0}(t,s),\\
	y_i(t)&=h_i(\psi_{t_0}^{u}(t,\gamma(s)),t), \quad i\in\{1,\cdots, n\}
	\end{split}
	\label{eq:controlsystemPath}	
	\end{equation}
	for all $t\geq t_0$, and all $s\in I$.The differential in direction $\frac{\partial}{\partial s}$ at a fixed   $t$ is a variation with respect to $s$. For the input-state-output solution $(u,x,y)$, the variations are
	\begin{equation}
	\delta u=\frac{\partial \varrho_{t_0}^{u}}{\partial s}(t,s); \  \delta x=\frac{\partial \psi_{t_0}^{u}}{\partial s}(t,\gamma(s));\  \delta y=\frac{\partial h}{\partial s}(\psi_{t_0}^{u},t),
	\end{equation}
	which are nothing, but tangent vectors to  $\varrho_{t_0}(t,s)$, $\psi_{t_0}^{u}(t,\gamma(s))$, and  $h_i(\psi_{t_0}^{u}(t,\gamma(s)),t))$, for $i\in\{1,\cdots, n\}$, respectively; i.e, $\delta u \in T_u\mathcal{U}$, $\delta x \in T_x\mathcal{X}$, and $\delta y \in T_y\mathcal{Y}$.  
	
	This let us to introduce the concept of prolongation to the tangent bundle of a given system  a long the the trajectory $(u,x,y)(t)=(\varrho_{t_0}(t,s),\psi_{t_0}^{u}(t,\gamma(s)),h_i(\psi_{t_0}^{u}(t,\gamma(s)),t)))$.
	
	\begin{definition}[\cite{crouch}]
		A \emph{ prolonged system}  of system \eqref{eq:controlsystem} corresponds to the original system \emph{together} with its \emph{variational system}, that is 
		\begin{equation}
		\begin{split}
		\left\{ \begin{array}{lcc}
		\dot{{x}}={f}({x},t)+\sum_{i=1}^{n}{g}_i({x},t){u}_i,\\
		y=h(x,t),\\
		\delta\dot{x}=\frac{\partial f}{\partial x}(x,t)\delta x+\sum_{i=1}^{n}\frac{\partial g}{\partial x}(x,t)\delta x+\sum_{i=1}^{n}g_i(x,t)\delta u,\\
		\delta y=\frac{\partial h}{\partial x}(x,t)\delta x.\\
		\end{array}
		\right.
		\end{split}
		\label{eq:prolongedcontrolsystem}
		\end{equation}
		with   $({u},\delta{u})\in T\mathcal{U}$,   $({x},\delta{x})\in T\mathcal{X}$, and  $({y},\delta{y})\in T\mathcal{Y}$.
	\end{definition}				
	
	System \eqref{eq:controlsystem} in closed-loop with the uniformly smooth static  feedback control law $u=\eta(x,t)$ will be denoted by 
	\begin{equation}
	\dot{{x}}={F}({x},t).
	\label{eq:closedloopcontrolsystem}
	\end{equation}
%
	
	%
	\subsection{Contraction and differential Lyapunov theory}

	\begin{definition}[\cite{forni}]\label{defin:finslerlyapunov}
		A function  $V:T\mathcal{X}\times \mathds{R}_{\geq 0}\rightarrow \mathds{R}_{\geq 0}$ is a candidate \emph{differential or Finsler-Lyapunov function} if it satisfies uniformly the bounds 
		\begin{equation}
		c_1 \mathcal{F}({x},\delta{x},t)^{p} \leq V({x},\delta{x},t) \leq c_2\mathcal{F}({x},\delta{x},t)^{p},
		\label{eq:finslerlyapunov}
		\end{equation}
		where $c_1,c_2\in\mathds{R}_{>0}$, $p$ is some positive integer and $\mathcal{F}({x},\cdot,t):=\|\cdot \|_{{x},t}$ is,  uniformly in $t$, a Finsler structure.
	\end{definition}
	The relation between a candidate differential Lyapunov function and the Finsler structure in \eqref{eq:finslerlyapunov} is a key property for incremental stability analysis. That is,  a well-defined distance on $\mathcal{X}$ via integration as defined below.
	\begin{definition}[Finsler distance]\label{defin:finslerdistance}
		Consider a candidate differential Lyapunov function on   $\mathcal{X}$ and the associated Finsler structure $\mathcal{F}$. Let $\Gamma({x}_1,{x}_{2})$ be the collection of piecewise $C^1$ curves ${\gamma}:I\rightarrow \mathcal{X}$ connecting ${x_1}$ and  ${x}_2$ such that ${\gamma}(0)={x}_1$ and ${\gamma}(1)={x}_2$. The Finsler distance $d:\mathcal{X}\times\mathcal{X}\rightarrow\mathds{R}_{\geq 0}$ induced by $\mathcal{F}$ is defined by
		\begin{equation}
		d({x}_1,{x}_2):=\inf_{\Gamma({x}_1,{x}_2)}\int_{\gamma}{\mathcal{F}\left({\gamma}(s),\frac{\partial {\gamma}}{\partial s}(s),t\right)}ds.
		\label{defin:finslerdistance1}
		\end{equation}
	\end{definition}
	
	The following result gives a sufficient condition for contraction by using differential Lyapunov functions. 
	\begin{theorem}[\cite{forni}]\label{theo:lyapunovcontraction}
		Consider the   prolonged system of \eqref{eq:closedloopcontrolsystem}, a connected and forward invariant set $\mathcal{D}\subseteq\mathcal{X}$, and a strictly increasing function $\alpha:\mathds{R}_{\geq 0}\rightarrow \mathds{R}_{\geq 0}$. Let $V$ be a candidate differential Lyapunov function satisfying
		\begin{equation}
		\dot{V}({x},\delta {x},t)	\leq -\alpha(V({x},\delta {x},t)) 
		\label{eq:finslerlyapunovinequality}
		\end{equation}
		for each $({x},\delta{x})\in T\mathcal{X}$ and uniformly in $t$. Then, system \eqref{eq:closedloopcontrolsystem} contracts $V$ in $\mathcal{D}$. The function $V$ is called the \emph{contraction measure}, and $\mathcal{D}$ is   \emph{the contraction region}. 
	\end{theorem}
	Contraction of \eqref{eq:closedloopcontrolsystem} is guaranteed by \eqref{eq:finslerlyapunov} and \eqref{eq:finslerlyapunovinequality}, with respect to the distance \eqref{defin:finslerdistance1}. Consequently, we have
	\begin{corollary}\label{remark:incremental}
		System \eqref{eq:closedloopcontrolsystem} is incrementally
		\begin{itemize}
			\item   \emph{stable} on $\mathcal{D}$ if $\alpha(s)=0$ for each $s\geq 0$;
			\item  \emph{asymptotically stable} on $\mathcal{D}$ if $\alpha$ is a strictly increasing;
			\item  \emph{exponentially stable} on $\mathcal{D}$ if $\alpha(s)=\beta s, \forall s>0.$
		\end{itemize}
	\end{corollary}
	
	\begin{remark}
		Under hypothesis of Theorem \ref{theo:lyapunovcontraction}, if the contraction region $\mathcal{D}\subseteq\mathcal{X}$ is a compact set. Then, system \eqref{eq:closedloopcontrolsystem} is  \emph{convergent}, \cite{ruffer2013convergent}. In this case, \eqref{eq:finslerlyapunovinequality} could be seen as a generalization of the Demidovich condition \cite{pavlov2017convergent}.
	\end{remark}	
	
	\subsection{Differential passivity}
	
	
	\begin{definition}[\cite{forni2013differential,arjan2013differentialpassivity}]
		Consider   system \eqref{eq:prolongedcontrolsystem}. Then, system \eqref{eq:controlsystem} is called \emph{differentially passive} if the prolonged system \eqref{eq:prolongedcontrolsystem}  is dissipative with respect to the supply rate $\delta y^{\top}\delta u$,  that is, if there exist a \emph{differential storage function} function ${W}:T\mathcal{X}\rightarrow\mathds{R}_{\geq 0} $ satisfying
		\begin{equation}
		\frac{d W}{dt}(x,\delta x)\leq \delta y^{\top}\delta u,
		\label{eq:differentialpassivityinequality}
		\end{equation}
		for all $x,\delta x, u,\delta u$. Furthermore,  system \eqref{eq:controlsystem} is called \emph{differentially loss-less} if \eqref{eq:differentialpassivityinequality} holds with equality.
	\end{definition}
	
	If additionally, the differential storage function is required to be a differential Lyapunov function, then differential passivity implies contraction when the input is $u_i=0$, for all $i\in\{1,\dots,n\}$. For further details and a differential geometric characterization see \cite{arjan2013differentialpassivity}. 
	
	The following result will be extensively used in the paper.
	\begin{lemma}\label{lemma:variationalpH-like}
		Consider  system \eqref{eq:controlsystem}.  Suppose that the control  $u=\eta(x,t)+\omega$ is designed   such that, via the differential transformation $\delta \tilde{x}=  \Theta(x,t)\delta x$, the variational dynamics  of the closed-loop system takes the form
		\begin{equation}
		\delta \dot{\tilde{x}}=\left[\Xi(\tilde{x},t)-\Upsilon(\tilde{x},t)\right]\Pi(\tilde{x},t)\delta\tilde{x}+\Psi(\tilde{x},t)\delta \omega,
		\label{eq:feedbackinterconnectionsystemsvariational}	
		\end{equation}
		where $\omega$ is an auxiliary input, $\Pi(\tilde{x},t)>0$ is a Riemannian metric, $\Xi(\tilde{x},t)=-\Xi^{\top}(\tilde{x},t)$ and  $\Upsilon(\tilde{x},t)=\Upsilon^{\top}(\tilde{x},t)$ satisfying $\delta\tilde{x}\left[\dot{\Pi}(\tilde{x},t)-2\Pi(\tilde{x},t)\Upsilon(\tilde{x},t)\Pi(\tilde{x},t)\right]\delta\tilde{x}\leq0$. Then, the closed-loop  system is differentially passive from $\delta\omega$ to $\delta\tilde{y}=\Psi(\tilde{x},t)^{\top}\Pi(\tilde{x},t)\delta \tilde{x}$ and differential storage function 
		\begin{equation}
		V(\tilde{x},\delta \tilde{x})=\frac{1}{2}\delta \tilde{x}^{\top}\overline{\Pi}(\tilde{x},t)\delta \tilde{x}.
		\label{eq:differentialStorage-Lemma}
		\end{equation}		
	\end{lemma}
%
%
%
	
	\subsection{Contraction and differential passivity of virtual systems}
	A generalization of contraction was first  introduced in \cite{wang} and revisited in \cite{jouffroy,forni},  with the name of \emph{partial contraction},  which is based on the  contraction behavior of the so-called  {virtual systems}.
	
	\begin{definition}\label{defin:virtualsystem1}
		A \emph{virtual system} associated to \eqref{eq:closedloopcontrolsystem} is defined as a system 
		\begin{equation}
		\dot{{x}}_v={\Phi}(x_v,{x},t),
		\label{eq:virtualsystemTheorem}
		\end{equation}
		in the state $x_v\in\mathcal{C}_v$ and  parametrized by $x\in\mathcal{C}_x$,  where $\mathcal{C}_v\subseteq \mathcal{X}$ and $\mathcal{C}_x\subseteq \mathcal{X}$ are connected and forward invariant,   $\Phi:\mathcal{X}\times\mathcal{X}\times\mathds{R}_{\geq 0}\rightarrow T\mathcal{X}$ is a smooth vector field satisfying 
		\begin{equation}
		{\Phi}({x},{x},t)=F(x,t),  \quad \forall t\geq t_0.
		\label{eq:virtualcontrolsystemTheorem}		
		\end{equation}
		Furthermore, a  \emph{virtual control system} for the system with inputs \eqref{eq:controlsystem},  in  the state $x_v\in\mathcal{X}$, is similarly defined as
		\begin{equation}
		\begin{split}
		\dot{x}_v&=\Gamma(x_v,x,u,t),\\
		y_v&=h_v(x_v,x,t)
		\end{split}
		\label{eq:virtualsystemTheorem1}	
		\end{equation}
		parametrized by the variable $x\in\mathcal{X}$, the output $y_v\in\mathcal{Y}$, with smooth vector fields $h_v:\mathcal{X}\times\mathcal{X}\times\mathds{R}_{\geq0}\rightarrow  \mathcal{Y}$ and  $\Gamma:\mathcal{X}\times\mathcal{X}\times\mathcal{U}\times\mathds{R}_{\geq0}\rightarrow T \mathcal{X}$ satisfying
		\begin{equation}
		\begin{split}
		\Gamma(x,x,u,t)&=f(x,t)+G(x,t)u,\\
		h_v(x,x,t)&=h(x,t),\quad\quad\quad \forall u, \forall t\geq t_0.		
		\end{split}
		\end{equation}		
	\end{definition}	
	It follows that any solution $x(t)=\psi_{t_0}(t,x_o)$ starting at $x_0\in\mathcal{C}_x$ of the \emph{actual system} \eqref{eq:closedloopcontrolsystem}, generates the solution $x_v(t)=\psi_{t_0}(t,x_o)$ to  system \eqref{eq:virtualsystemTheorem} for all $t>t_0$. In a similar manner, any solution of  \eqref{eq:controlsystem} $x=\psi_{t_0}^{u}(t,x_0)$, for a certain input $u=\tau\in\mathcal{U}$, generates a solution $x_v(t)=\psi_{t_0}^{u}(t,x_0)$ to the {virtual control system} \eqref{eq:virtualsystemTheorem1}. However. \emph{not} every solution $x_v(t)$ of the virtual system, corresponds to a solution of the actual system. Thus, \emph{for \emph{any} curve $x(t)$, we may consider the time-varying system  with state $x_v$.}	

	The convergence behavior of  \eqref{eq:closedloopcontrolsystem} can be induced from the contraction properties of an associated virtual system \cite{wang,forni}.
	
	\begin{theorem}\label{theo:partialcontraction}
		Consider two connected and forward invariant sets $\mathcal{C}_x\subseteq \mathcal{X}$  and $\mathcal{C}_v\subseteq \mathcal{X}$ for  systems \eqref{eq:closedloopcontrolsystem} and \eqref{eq:virtualsystemTheorem} respectively. Suppose that system \eqref{eq:virtualsystemTheorem}  is uniformly contracting with respect to $x_v$. Then, for any given initial conditions $x_0\in\mathcal{C}_x$, and $x_{v0}\in\mathcal{C}_v$, each solution to \eqref{eq:virtualsystemTheorem} converges asymptotically to the solution  of \eqref{eq:closedloopcontrolsystem}.
	\end{theorem}	
%
%
	If this holds, the actual system  \eqref{eq:closedloopcontrolsystem} is said to be \emph{virtually contracting} to the virtual system  \eqref{eq:virtualsystemTheorem}. This does not imply that the actual system is contracting, but all its trajectories converge to the steady state solution of the virtual system. Thus,  \emph{if a system is virtually contracting, then it is convergent} as in  \cite{pavlov2017convergent}.

	Furthermore, if there exist a virtual control system for \eqref{eq:controlsystem}, such that it is differentially passive, then the  actual control system is said to be \emph{virtually differentially passive}. 
	\subsection{Virtual differential passivity based control}\label{subsection:v-dPBC}				
	We propose a constructive  control design method for system \eqref{eq:controlsystem},  that we shall call \emph{virtual differential passivity based control (v-dPBC)}, such that the closed-loop system is  \emph{convergent} to a desired behavior.  The  design procedure  is divided in three main steps:
	\begin{enumerate}
		\item Design the virtual control system  \eqref{eq:virtualsystemTheorem1} for system  \eqref{eq:controlsystem}.
		\item Design the feedback $u=\eta(x_v,x,t)+\omega$ for \eqref{eq:virtualsystemTheorem1} such that the closed-loop virtual system   is differentially  passive for  the input-output pair  $(\delta {{y}}_v,\omega)$ and has a desired trajectory $x_d(t)$ as steady-state   solution.
		\item Define the controller for  system \eqref{eq:controlsystem} as $u=\eta(x,x,t)$.
	\end{enumerate}
	
	If we are able to design a controller following the above steps, then we   all   closed-loop system  trajectories will  converge to $x_d(t)$, for the  external input $\omega=0$.

\section{Mechanical port-Hamiltonian systems}	
	Ideas in the previous section will be applied mechanical systems in the pH framework, which are described below.
	\begin{definition}[\cite{arjanl2}]
		A {port-Hamiltonian system} with a $N$-dimensional state space manifold $\mathcal{X}$, input and output spaces $\mathcal{U}=\mathcal{Y}\subset\mathds{R}^{m}$, and Hamiltonian function $H:\mathcal{X}\rightarrow\mathds{R}$, is given by
		\begin{equation}
		\begin{split}
		\dot{x}&=\left[J(x)-R(x)\right]\frac{\partial H}{\partial x}(x)+g(x)u\\			
		y&= g^{\top}(x)\frac{\partial H}{\partial x}(x),
		\end{split}
		\label{eq:IOpHsystem}			
		\end{equation}
		where $g(x)$ is a $N\times m$ input matrix, and   $J(x)$, $R(x)$ are the interconnection and dissipation $N\times N$ matrices which satisfy $J(x)=-J^{\top}(x)$ and $R(x)=R^{\top}(x)\geq0$. 
	\end{definition}
	
	In the  case of a standard mechanical system with generalized coordinates $q$ on the configuration space $\mathcal{Q}$ of dimension $n$ and  velocity $\dot{{q}}\in T_{q}\mathcal{Q}$, the Hamiltonian function is given by the total energy
	\begin{equation}
	H({x})=\frac{1}{2}{p}^{\top}{M}^{-1}({q}){p}+P({q}),
	\label{eq:phmechanicalHamiltonian}
	\end{equation}
	where ${x}=({q},{p})\in T^*\mathcal{Q}:=\mathcal{X}$ is the phase state, $P({q})$ is the potential energy,  $p:=M(q)\dot{q}$ is the generalized momentum and the inertia matrix $M(q)$ is symmetric and positive definitive; Finally, the interconnection, dissipation and input matrices in \eqref{eq:IOpHsystem} are
	\begin{equation}
	J(x)=\begin{bmatrix}
	{0}_n & {I}_n\\
	-{I}_n& {0}_n
	\end{bmatrix};  R(x)=\begin{bmatrix}
	{0}_n & {0}_n\\
	{0}_n& {D}({{q}})
	\end{bmatrix}; g(x)=\begin{bmatrix}
	{0}_n\\
	B(q)
	\end{bmatrix},
	\label{eq:phmechanical}		
	\end{equation}		
	with matrix  ${D}({q})={D}^{\top}({q}) \geq {0}_n$ being the damping matrix and  ${I}_n $, ${0}_n$ the $n\times n$ identity, respectively, zero matrices. The input force matrix $B(q)$ has rank $m\leq n$.

	In \cite{Arimoto}, it was shown that the identity
	\begin{equation}
	\frac{1}{2}\dot{q}^{\top}\dot{M}(q)\dot{q}=\dot{q}^{\top}\frac{\partial}{\partial q}\left(\frac{1}{2}\dot{q}^{\top}M(q)\dot{q}\right)
	\label{eq:SuguruProperty}
	\end{equation}
	implies the existence of a (\emph{gyroscopic forces}) skew-symmetric matrix $S_
		L(q,\dot{q})$ that satisfies the relation		
		\begin{equation}
		-\frac{\partial}{\partial {q}}\left[\frac{1}{2}\dot{{q}}^{\top}{M}({q})\dot{{q}}\right]=\left[{S}_L({q},{\dot{q}})-\frac{1}{2}\dot{{M}}({q})\right]{\dot{q}}.
		\label{eq:coriolislagrange}
		\end{equation}			
In order to express the relation \eqref{eq:coriolislagrange}				 in the Hamiltonian framework, consider the generalized momentum and the Legendre transformation of the quadratic form in the brackets of the left hand side (\cite{arjanl2}). Then, the following holds
	\begin{equation}
	\frac{\partial}{\partial q}\left(\frac{1}{2}p^{\top}M^{-1}(q)p\right)=-\frac{\partial}{\partial q}\left(\frac{1}{2}\dot{q}^{\top}M(q)\dot{q}\right).
	\end{equation}
	With this, \eqref{eq:coriolislagrange} can be rewritten  in terms of $(q,p)$ as
	\begin{equation}
	\frac{\partial }{\partial {q}}\left[\frac{1}{2}{{p}}^{\top}{M}^{-1}({q}){{p}}\right]=\left[S_H(q,p)-\frac{1}{2}\dot{{M}}({q})\right]{M}^{-1}({q}){p}.
	\label{eq:coriolishamilton}
	\end{equation}
	where the matrix $S_H(x):=M(q){\overline{S}}_H(x)M(q)$, which is nothing but $S_L(q,\dot{q})$ in coordinates  $x\in T^{*}\mathcal{X}$. 
	
	Then,  system \eqref{eq:IOpHsystem}-\eqref{eq:phmechanical}  can be rewritten as
	\begin{equation}
	\begin{split}
	\begin{bmatrix}
	\dot{{q}}\\
	\dot{{p}}
	\end{bmatrix}&=\begin{bmatrix}
	{0}_n & {I}_n\\
	-{I}_n& -({E}(q,p)+D(q))
	\end{bmatrix}\begin{bmatrix}
	\frac{\partial P}{\partial {q}}({q})\hfill \\
	\frac{\partial H}{\partial {p}}(q,p)
	\end{bmatrix}+g(q){u},\\
	y&=\begin{bmatrix}
	{0}_n& {B}^{\top}({q})
	\end{bmatrix}\begin{bmatrix}
	\frac{\partial P}{\partial {q}}({q})\hfill \\
	\frac{\partial H}{\partial {p}}(q,p)
	\end{bmatrix},
	\end{split}
	\label{eq:phmechanica3}
	\end{equation}
	with 
	\begin{equation}
	{E}(q,p):={S}_H(q,p)-\frac{1}{2}\dot{{M}}({q}).
	\label{eq:worklessforcesmatrix}
	\end{equation}
	The structure of the matrix $E(q,p)$ and the conservation of energy  tell us that the associated forces  are \emph{workless}. 	

	Based on the above description, we can define a pH-like virtual control system as in \eqref{eq:virtualsystemTheorem1} associated to system \eqref{eq:IOpHsystem}-\eqref{eq:phmechanical},  with the state $x_v=(q_v, p_v)\in \mathcal{X}$ and  parametrized by   $x=(q,p)$, as follows
	\begin{equation}
	\begin{split}
	\dot{x}_v&=\left[J_v(x)-R_v(x)\right]\frac{\partial H_v}{\partial x_v}(x_v,x)+g(x){u},\\
		y_v&=g^{\top}(x)\frac{\partial H_v}{\partial x_v}(x_v,x),
	\end{split}
	\label{eq:phmechanicalvirtual}
	\end{equation}
	with matrices  $J_v=-J^{\top}_v$ and $R_v=R_v^{\top}$ defined by
	\begin{equation}
	J_v=\begin{bmatrix}
	0_n & I_n\\
	-I_n & -S_H
	\end{bmatrix}, \quad R_v:=\begin{bmatrix}
	0_n & 0_n\\
	0_n & (D-\frac{1}{2}\dot{M})
	\end{bmatrix}.
	\label{eq:pH-likesystem}
	\end{equation}		
	and	 virtual Hamiltonian function
	\begin{equation}
	H_v(x_v,x)=\frac{1}{2}p_v^{\top}M^{-1}(q)p_v+P(q_v).
	\label{eq:phmechanicalHamiltonianvirtual}
	\end{equation}

	\begin{remark}
		Matrix $J_v(x)$ qualifies as  interconnection structure of the virtual system \eqref{eq:phmechanicalvirtual}. However, the matrix $R_v(x)$  is not necessarily positive definite. This is the reason why the  virtual system \eqref{eq:phmechanicalvirtual} is called a \emph{pH-like system}.
	\end{remark}	
	

	The variational virtual system of \eqref{eq:phmechanicalvirtual} is given by
	\begin{equation}
	\begin{split}
	\delta \dot{x}_v&=\left[J_v(x)-R_v(x)\right]\frac{\partial^2 H_v}{\partial x_v^2}(x_v,x)\delta x_v+g(x)\delta u		\\
	\delta y_v&=g^{\top}(x)\frac{\partial^2 H_v}{\partial x_v^2}(x_v,x)\delta x_v.
	\end{split}
	\label{eq:variationalvirtualpHsystem}
	\end{equation}

    System  \eqref{eq:variationalvirtualpHsystem} has the form   \eqref{eq:feedbackinterconnectionsystemsvariational} with $\Xi=J_v$, $\Upsilon=R_v$ and $\Pi(x_v,x)=\frac{\partial^2 H_v}{\partial x_v^2}(x_v,x)$.   If $\frac{\partial ^2 P}{\partial q_v^2}>0$ is such that hypotheses in Lemma \ref{lemma:variationalpH-like} are satisfied, then the  pH-like virtual system \eqref{eq:phmechanicalvirtual} is \emph{differentially passive} with differential storage function
	\begin{equation}
	V(x_v,\delta x_v,x)=\frac{1}{2}\delta x_v^{\top}\Pi(x_v,x)\delta x_v.
	\end{equation}	

	 The   presents a  controller for fully-actuated system using v-dPBC i.e.,  $n=m$. This controller will be used in next section.  For notational purposes  we add the subscript $\ell$ in all the terms that define  systems  \eqref{eq:phmechanical} and \eqref{eq:phmechanicalvirtual}, e.g., $x=x_{\ell}$, $x_v=x_{\ell v}$, $u=u_{\ell}$, etc.

	\begin{lemma}[\cite{RodoAutomatica2017}]\label{lemma:fullstatecontroller-rigid}
		Consider a desired smooth  trajectory ${x}_{\ell d}=(q_{\ell d},p_{\ell d})\in T^*\mathcal{Q}_{\ell}$, with $n_{\ell}=\text{dim}\mathcal{Q}_{\ell}$. Let us introduce the following  change coordinates
		\begin{equation}
		\tilde{{x}}_{\ell v}:=\begin{bmatrix}
		\tilde{{q}}_{\ell v}\\
		{\sigma_{ \ell z}}
		\end{bmatrix}=\begin{bmatrix}
		{q}_{\ell v}-{q}_{\ell d}\\
		{p}_{\ell v}-{p}_{\ell r}
		\end{bmatrix},
		\label{eq:changeofcoordinatesError-rigid}
		\end{equation}
		and define the auxiliary momentum reference as
		\begin{equation}
		p_{\ell r}:=M_{\ell}(q_{\ell})(\dot{q}_{\ell d}-\phi_{\ell}(\tilde{q}_{\ell v})),
		\label{eq:auxiliarreference-rigid}
		\end{equation}
		with $\phi_{\ell}:\mathcal{Q}\rightarrow T_{\tilde{q}_v}\mathcal{Q}_{\ell}$  and a positive definite Riemannian metric $\Pi_{\ell}:\mathcal{Q}_{\ell}\times \mathds{R}_{\geq 0}\rightarrow \mathds{R}^{n_{\ell}\times n_{\ell}}$  satisfying the inequality
		\begin{equation}
		\begin{split}
		\dot{\Pi}_{\ell}(\tilde{q}_{\ell v},t)-&\Pi_{\ell }(\tilde{q}_{\ell v},t)\frac{\partial \phi_{\ell}}{\partial \tilde{q}_{\ell v}}(\tilde{q}_{\ell v})-\frac{\partial \phi_{\ell}^{\top}}{\partial \tilde{q}_{\ell v}}(\tilde{q}_{\ell v})\\&\times\Pi_{\ell}(\tilde{q}_{\ell v},t)\leq  -2\beta_{\ell}(\tilde{q}_{\ell v},t) \Pi_{\ell}(\tilde{q}_{\ell v},t),
		\end{split}
		\label{eq:controllaw-positionmetricinquality-rigid}
		\end{equation}							
		with $\beta_{\ell}(\tilde{q}_{\ell v},t)>0$, uniformly. Consider also  system \eqref{eq:phmechanical}, its virtual system \eqref{eq:phmechanicalvirtual} and the composite control law given by  ${u}_{\ell}(x_{\ell v},x_{\ell},t):={u}_{\ell ff}+{u}_{\ell fb}$ with
		\begin{equation}
		\begin{split}
		{u}_{\ell ff}&=\dot{p}_{\ell r}+\frac{\partial P_{\ell}}{\partial q_{\ell}}+\big[E_{\ell} +D_{\ell}\big] M^{-1}_{\ell}(q(t))p_{\ell r},\\
		{u}_{\ell fb}&=-\int_{0_{n_{\ell}}}^{\tilde{{q}_{\ell v}}}\Pi_{\ell}(\overline{q}_{\ell v})\text{d}\overline{q}_{\ell v}-{K}_{\ell d}M^{-1}_{\ell}\sigma_{\ell v}+\omega_{\ell},
		\end{split}
		\label{eq:controlaw-rigid}
		\end{equation}
		where ${K}_{\ell d}>0$ and $\omega_{\ell}$ is an external input. Then, virtual system \eqref{eq:phmechanicalvirtual} in closed-loop with \eqref{eq:controlaw-rigid}  is differentially passive for the input-output pair $(\delta \omega,\delta y_{\sigma_{\ell v}})$, with $\delta y_{\sigma_{\ell v}}=B_{\ell}^{\top}M_{\ell}^{-1}\delta\sigma_{\ell v}$  and  differential storage function
		\begin{equation}
		V_{\ell}(\tilde{x}_{\ell v},\delta \tilde{x}_{\ell v},t)=\frac{1}{2}\delta \tilde{x}_{\ell v}^{\top}\begin{bmatrix}
		\Pi_{\ell }(\tilde{q}_{\ell v},t)& 0_{n_\ell}\\
		0_{n_{\ell}} & M_{\ell}^{-1}
		\end{bmatrix}\delta \tilde{x}_{\ell v}.
		\label{eq:design-dLCF-sigma-rigid}
		\end{equation}				
	\end{lemma}

\section{Trajectory tracking controller of flexible-joint robots}	
\subsection{Flexible-joints Robots as port-Hamiltonian systems}

Flexible rotational joints robots are a particular class of  mechanical  systems \eqref{eq:phmechanical}, where the generalized position is split as $q=[q_{\ell}^{\top},q_m^{\top}]^{\top}\in\mathcal{Q}=\mathcal{Q}_{n_{\ell}}\times \mathcal{Q}_{n_m}$, where  $q_{\ell}$  and $q_m$ are the $n_{\ell}-$ links  and the $n_m-$motors generalized positions, respectively; with   $\text{dim}\mathcal{Q}=n_{\ell}+n_m$. The inertia and damping matrices are partitioned into   $M(q)=\text{diag}\{M_{\ell}(q_{\ell}),M_m(q_m)\}$ and ${D}({q})=\text{diag}\{D_{\ell}(q_{\ell}),D_m(q_{m})\} $, where $M_{\ell}(q_{\ell})$ and $M_m(q_m)$ are the link and motors inertias; similarly $D_{\ell}(q_{\ell})$ and $D_m(q_m)$ are the link and motor damping matrices, respectively. The potential energy    is
\begin{equation}
P(q)=P_{\ell}(q_{\ell})+\underbrace{\frac{1}{2}\zeta^{\top}K\zeta}_{P_m(q)},
\end{equation}
which is   the sum of the links potential energy $P_{\ell}(q_{\ell})$ and the joints potential energy $P_m(q)$, with $\zeta:=q_m-q_{\ell}$ and $K\in\mathds{R}^{n\times n}$ a symmetric, positive definitive matrix of stiffness coefficients. The  input acts only  in the motor state,  meaning that the flexible-joints robot is an \emph{underactuated system} and $\text{rank}(B(q))=n_m$. 
\begin{assumption}[ \cite{spongflexible}]
	The following standard assumptions on the system physical structure are made:
	\begin{itemize}
		\item The relative displacement $\zeta$ (deflection) at each joint is small, such that the spring's dynamics is linear. 
		\item The $i-$th motor, which drives the $i-th$ link, is mounted
		at the $(i−-1)$-th link.
		\item The center of mass of each motor is on the rotation axes.
		\item The angular velocity of the motors is due only to their own spinning.
	\end{itemize}	
\end{assumption}

Thus, a flexible-joints robot can be modeled as an underactuated pH system of the form \eqref{eq:phmechanical}, given by
\begin{equation}
\begin{split}
\begin{bmatrix}
\dot q_{\ell}\\
\dot q_m\\
\dot p_{\ell}\\
\dot p_m
\end{bmatrix} &= \begin{bmatrix}
0_{n_{\ell}} & 0_{n_m} & I_{n_{\ell}} & 0_{n_m} \\
0_{n_{\ell}} & 0_{n_m}  & 0_{n_{\ell}} & I_{n_m} \\
-I_{n_{\ell}} & 0_m & -D_{\ell} & 0_{n_m} \\
0_{n_{\ell}} & -I_{n_m}  & 0_{n_{\ell}} & -D_m
\end{bmatrix}\frac{\partial H}{\partial x} + \begin{bmatrix}
0_{n_{\ell}}\\
0_{n_m} \\
0_{n_{\ell}}\\
B_m(q_m)
\end{bmatrix}u,\\
y&=B_m(q_m)^{\top}\frac{\partial H}{\partial p_m}(x),
\end{split}
\label{eq:phmechanical-flexible}
\end{equation}
where $p_{\ell}$ and $p_m$ are the  links and motors momenta, $p=[p^{\top}_{\ell},p^{\top}_m]^{\top}$ and $B_m(q_m)$ is the input matrix associated to the motors. System \eqref{eq:phmechanical-flexible} can be rewritten  as \eqref{eq:phmechanica3}, with 
\begin{equation}
E(x)=\begin{bmatrix}
S_{\ell}(q_{\ell},p_{\ell}) -\frac{1}{2}\dot{M}_{\ell}& 0_{n_m}\\
0_{n_{\ell}} & S_{m}(q_{m},p_{m})-\frac{1}{2}\dot{M}_m 
\end{bmatrix},
\label{eq:workless-flexible}
\end{equation}
with $S_{\ell}^{\top}=-S_{\ell}$ and $S_{m}^{\top}=-S_{m}$. With this specification, the virtual system \eqref{eq:phmechanicalvirtual} corresponding to \eqref{eq:phmechanical-flexible} is  
\begin{equation}
\begin{split}
\dot{x}_v &=\begin{bmatrix}
0_n & 0_n & I_n & 0_n\\
0_n & 0_n & 0_n & I_n\\
-I_n & 0_n & -(E_{11}+D_{\ell} )& 0_n\\
0_n & -I_n & 0_n & -(E_{22}+D_m )
\end{bmatrix}\frac{\partial H_v}{\partial x_v}+g(x){u},\\
y_v&=g^{\top}(x)\frac{\partial H_v}{\partial x_v}(x_v,t).
\end{split}
\label{eq:phmechanical-flexible-virtual}
\end{equation}

\subsection{Tracking controller design}

In this section we extent the controller of Lemma \ref{lemma:fullstatecontroller-rigid} to the underactuated  pH system \eqref{eq:phmechanical-flexible},   using the v-dPBC technique described in Subsection \ref{subsection:v-dPBC} with respect to the virtual system \eqref{eq:phmechanical-flexible-virtual} .  Through	a recursive	construction of \emph{{differential} storage functions}, we will implicitly design the differential transformation $\Theta(x_v,t)$ such that the closed-loop variational virtual system  satisfies Lemma \ref{lemma:variationalpH-like}. 

	\begin{proposition}\label{proposition:fullstatecontroller-flexible}
		Consider the virtual system of FJRs in\eqref{eq:phmechanical-flexible-virtual}.  Suppose that the hypotheses  and controller in Lemma \ref{lemma:fullstatecontroller-rigid} hold for the link dynamics with the controller $u_{\ell}$ given by \eqref{eq:controlaw-rigid}. Let the  motor  reference state be given by ${x}_{m d}=(q_{m d},p_{m d})\in T^*\mathcal{Q}_m$, with $q_{md}=q_{\ell}+K^{-1}u_{\ell}$ and $n_{m}=\text{dim}\mathcal{Q}_m$. Consider   the following change of coordinates
		\begin{equation}
		\tilde{{x}}_{mv}:=\begin{bmatrix}
		\tilde{{q}}_{mv}\\
		{\sigma_{mv}}
		\end{bmatrix}=\begin{bmatrix}
		{q}_{mv}-{q}_{md}\\
		{p}_{mv}-{p}_{mr}
		\end{bmatrix},
		\label{eq:changeofcoordinatesError-motor}
		\end{equation}
		and define the auxiliary motor momentum reference as
		\begin{equation}
		p_{mr}:=M_{m}(q_{m}(t))(\dot{q}_{md}-\phi_{m}(\tilde{q}_v)-\Pi^{-1}_{m}K^{\top}M_{\ell}^{\top}\sigma_{\ell v}),
		\label{eq:auxiliarreference-motor}
		\end{equation}
		where $\phi_{m}:\mathcal{Q}_m\rightarrow T_{\tilde{q}_v}\mathcal{Q}_m$  and a positive definite Riemannian metric $\Pi_{m}:\mathcal{Q}_m\times \mathds{R}_{\geq 0}\rightarrow \mathds{R}^{n_m\times n_m}$  satisfying  
		\begin{equation}
		\begin{split}
		\dot{\Pi}_{m}(\tilde{q}_{m v},t)-&\Pi_{m }(\tilde{q}_{m v},t)\frac{\partial \phi_{m}}{\partial \tilde{q}_{m v}}(\tilde{q}_{m v})-\frac{\partial \phi_{m}^{\top}}{\partial \tilde{q}_{m v}}(\tilde{q}_{m v})\\&\times\Pi_{m}(\tilde{q}_{m v},t)\leq  -2\beta_{m}(\tilde{q}_{m v},t) \Pi_{m}(\tilde{q}_{m v},t),
		\end{split}
		\label{eq:controllaw-positionmetricinquality-motor}
		\end{equation}							
		with $\beta_{m}(\tilde{q}_{m v},t)>0$. Consider also  system \eqref{eq:phmechanical-flexible}, its corresponding virtual system \eqref{eq:phmechanical-flexible-virtual} and the  control law given by  ${u}(x_{z},x,t):={u}_{mff}+{u}_{mfb}$ with
		\begin{equation}
		\begin{split}
		{u}_{mff}&=\dot{p}_{mr}+k\zeta +\big[E_m+D_m\big] M_m^{-1}(q_m)p_{mr},\\
		{u}_{mfb}&=-\int_{0_{n_m}}^{\tilde{{q}_{mv}}}\Pi_{m}(\overline{q}_{mv})\text{d}\overline{q}_{mv}-{K}_{md}M^{-1}_m\sigma_{mv}+\omega,
		\end{split}
		\label{eq:controlaw-motor}
		\end{equation}
		where ${K}_{md}>0$ and $\omega$ and external input. Then,  the closed-loop virtual system \eqref{eq:phmechanical-flexible-virtual}  is differentially passive with respect to input-output pair $(\delta \omega,\delta y_{\sigma_{mv}})$, with $\delta y_{\sigma_{mv}}=B_m^{\top}M_m^{-1}\delta\sigma_{mv}$  and   differential storage function
			\begin{equation}
			V(\tilde{x}_v,\delta \tilde{x}_v,t)=\frac{1}{2}\delta \tilde{x}_v^{\top}\begin{bmatrix}
			\Pi(\tilde{q}_v,t)& 0_n\\
			0_n & M^{-1}(q)
			\end{bmatrix}\delta \tilde{x}_v,
			\label{eq:design-dLCF-sigma-flexible}
			\end{equation}		
		with $\Pi=\text{diag}\{\Pi_{\ell}(\tilde{q}_{\ell v}),\Pi_m(\tilde{q}_{mv})\}$.	Furthermore, the closed-loop variational dynamics of \eqref{eq:phmechanical-flexible-virtual} preserves the  structure of \eqref{eq:variationalvirtualpHsystem} with  
		\begin{equation}
		\begin{split}
		\frac{\partial^2 H_v}{\partial \tilde{x}_v^2}&=\text{diag}\big\{\Pi, M^{-1}(q)\big\},\\
		R_v(\tilde{x})&=\text{diag}\bigg\{\frac{\partial \phi}{\partial \tilde{q}_v}\Pi^{-1},[E(x)+D(q)+K_d] \bigg\},\\
		J_v(\tilde{x})&=\begin{bmatrix}
		0_{n_{\ell}} & 0_{n_{m}} & I_{n_{\ell}} & 0_{n_m}\\
		0_{n_{\ell}} & 0_{n_{m}} & -\Pi_m^{-1}K^{\top} &  I_{n_{m}}\\
		 -I_{n_{\ell}} & K\Pi_m^{-1} & 0_{n_{\ell}} &  0_{n_{m}}\\
		 0_{n_{\ell}}  &  - I_{n_{m}} & 0_{n_{\ell}} &  0_{n_m}
		\end{bmatrix},
		\end{split}
		\label{eq:feedbackinterconnectionsystemsvariational-flexible}
		\end{equation}
		where	$\phi(\tilde{q}_v)=[\phi^{\top}_{\ell},\phi^{\top}_m]^{\top}$ and $K_d=\text{diag}\{K_{\ell},K_m\}$.
	\end{proposition}
	Notice that  from \eqref{eq:feedbackinterconnectionsystemsvariational-flexible}, the closed-loop variational dynamics of the differentially passive system \eqref{eq:phmechanical-flexible-virtual} can be seen as the  feedback interconnection between the  variational virtual pH-like link error dynamics
	\begin{equation}
	\begin{split}
	\delta\dot{x}_{\ell v}&=\begin{bmatrix}
	-\frac{\partial \phi_{\ell}}{\partial \tilde{q}_{\ell z }}\Pi_{\ell v}^{-1} & I_{n_{\ell}}\\
	-I_{n_{\ell}} & -(E_{\ell}+D_{\ell}+K_{\ell d})
	\end{bmatrix}\frac{\partial V_{\ell} }{\partial \delta x_{\ell v}}+\begin{bmatrix}
	0_{n_{\ell}}\\
	\delta\overline{u}_{\ell 2}
	\end{bmatrix},\\
	\delta y_{\ell z }&=\begin{bmatrix}
	0_{n_{\ell}}  & 0_{n_{\ell}}\\
	0_{n_{\ell}} & I _{n_{\ell}}
	\end{bmatrix}\frac{\partial V_{\ell} }{\partial \delta x_{\ell v}},
	\end{split}
	\end{equation}
	with "differential Hamiltonian" function
	\begin{equation}
	 V_{\ell}=\frac{1}{2}\delta \tilde{x}_{\ell v}^{\top}\begin{bmatrix}
	 \Pi_{\ell}& 0_n\\
	 0_n & M^{-1}_{\ell}
	 \end{bmatrix}\delta \tilde{x}_{\ell v},
	 \label{eq:differentionalhamiltonianfunction-link}
	\end{equation}
	and the   variational pH-like motor link error dynamics
	\begin{equation}
	\begin{split}
	\delta\dot{x}_{m z}&=\begin{bmatrix}
	-\frac{\partial \phi_{m}}{\partial \tilde{q}_{m v }}\Pi_{m z}^{-1} & I_{n_{m}}\\
	-I_{n_{m}} & -(E_{m}+D_{m}+K_{m d})
	\end{bmatrix}\frac{\partial V_{m} }{\partial \delta x_{m v}}\\& +\begin{bmatrix}
	I_{n_m}  & 0_{n_m}\\
	0_{n_m} & I _{n_m}
	\end{bmatrix}\begin{bmatrix}
	\delta\overline{u}_{m1}\\
	\delta\overline{u}_{m2}
	\end{bmatrix},\\
	\delta y_{m z}&=\begin{bmatrix}
	I_{n_m}  & 0_{n_m}\\
	0_{n_m} & I _{n_m}
	\end{bmatrix}\frac{\partial V_{m} }{\partial \delta x_{m v}},
	\end{split}
	\end{equation}
	with "differential Hamiltonian" function
	\begin{equation}
	V_{m}=\frac{1}{2}\delta \tilde{x}_{m v}^{\top}\begin{bmatrix}
	\Pi_{m}& 0_n\\
	0_n & M^{-1}_{m}
	\end{bmatrix}\delta \tilde{x}_{m v},
	\label{eq:differentionalhamiltonianfunction-motor}
	\end{equation}		
   through the interconnection law
	\begin{equation}
	\begin{bmatrix}
	\delta \overline{u}_{\ell}\\
	\delta \overline{u}_{m}		
	\end{bmatrix}=\begin{bmatrix}
	0_{n_{\ell}}& 0_{n_{\ell}} & 0_{n_{m}} & 0_{n_{m}}\\
	0_{n_{\ell}}& 0_{n_{\ell}} &  K\Pi_{m} & 0_{n_{m}}\\
	0_{n_{\ell}}&-\Pi_{m}K^{\top} & 0_{n_{m}} & 0_{n_{m}}\\
	0_{n_{\ell}}& 0_{n_{\ell}} & 0_{n_{m}} & 0_{n_{m}}\\
	\end{bmatrix}\begin{bmatrix}
	\delta y_{{\ell}v}\\ 
	\delta y_{{m}v}
	\end{bmatrix}+\begin{bmatrix}
	0_{n_{\ell}} \\ 0_{n_{m}} \\ 0_{n_{\ell}} \\ I_{n_{m}}
	\end{bmatrix}\delta \omega.
	\end{equation}
	Hence, the {feedback interconnection} between the differentially passive links and motor error dynamics is a differentially passive closed-loop system with differential storage function $V=V_{\ell}+V_m$, as proved in \cite{arjan2013differentialpassivity}.

	\begin{corollary}[Trajectory tracking controller]
		Consider the controller \eqref{eq:controlaw-motor}. Then, all solutions of the flexible-joints robot \eqref{eq:phmechanical-flexible} in closed-loop with the controller $u(x,x,t)$ converges exponentially to the desired trajectory $x_d(t)$  with rate 
		\begin{equation}
		\begin{split}
		\beta&=\min\{\min\{\beta_{\ell},\beta_{m}\},\lambda_{\min}\{D+K_d\} \lambda_{\min}\{M^{-1}\}\}.
		\end{split}
		\end{equation}		
	\end{corollary}	
	
\section{Example: A Flexible-joint robot}	
	In this numerical example, we consider the FJR with one flexible joint i.e.,  $n_{\ell}=n_m=1$ \eqref{eq:phmechanical-flexible}. For the simulation, the parameters of the system and the controller as in \eqref{eq:controlaw-rigid} and $\eqref{eq:controlaw-motor}$ are given  in Table \ref{table:parametersexample}. Here, we consider the same setting of FJR as the one used in \cite{ghorbel1989adaptive}
\begin{table}[h!]
	\centering
	\begin{tabular}{ |l|l| }
		\hline
	{\bf	Parameter} & {\bf Value}\\
		\hline
		Link inertia, $M_{\ell}$ &  0.031 $kg\cdot m^2$ \\ 
		Rotor inertia, $M_{m}$ &  0.004 $kg\cdot m^2$ \\ 
		Rotor friction, $D_{\ell}$&   0.2 $N\cdot m\cdot sec/rad$ \\ 		
		Rotor friction, $D_{m}$&   0.007 $N\cdot m\cdot sec/rad$ \\ 
		Nominal load, $M_{\ell}g l$&   0.8 $N\cdot m$ \\ 		
		\hline	
		\textbf{Controller $u_{\ell}$} &\textbf{Controller $u_{m}$}\\
		\hline
		$\phi_{\ell}=\Lambda_{\ell} \tilde{q}_{\ell}$ &  $\phi_{m}=\Lambda_{m} \tilde{q}_{m}$ \\ 
		$\Lambda_{\ell} = 10$ & $\Lambda_{\ell} = 15$ \\ 
		$K_{\ell d} = 0.6$& 	$K_{\ell d} = 0.3$ \\ 		
		$\Pi_{\ell}= 2\Lambda_{\ell}$&  $\Pi_{\ell}= 4\Lambda_{m}$ \\ 
		\hline		
	\end{tabular}
	\caption{flexible joint link  parameters and control laws specifications}			
	\label{table:parametersexample}	
\end{table}

\begin{figure}[h!]
	\vspace{-0.2cm}	  	
	\includegraphics[width=0.48 \textwidth]{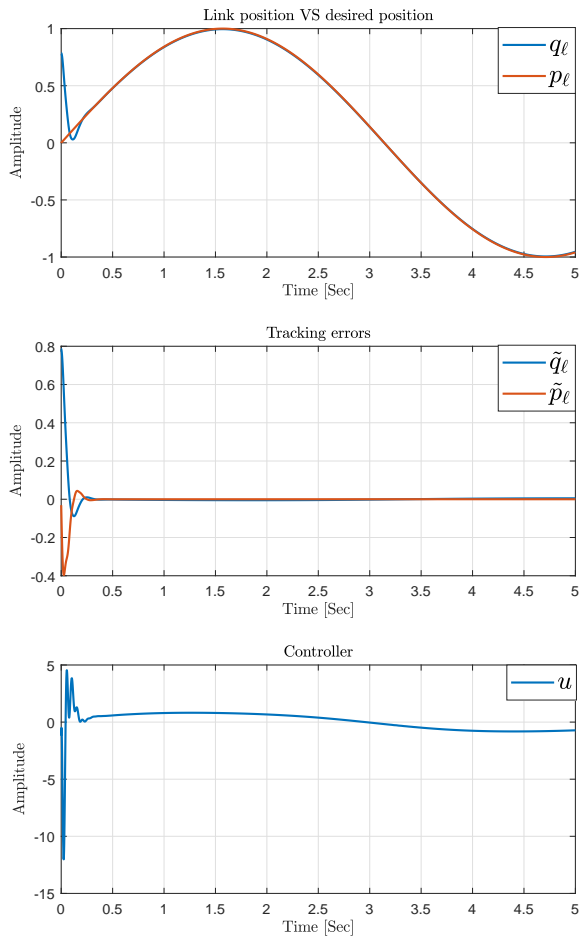}\hfill
	\vspace{-1cm}
	\caption{Performance with joint stiffness $k=31$  Nm/rad.}
	\label{fig:plots1k31}	
\end{figure}

The closed-loop performance  for the stiffness constant $k=31$ is shown in Figure \ref{fig:plots1k31}. After a short transient time, both, the position and momentum error converge to zero. The overshoot in the controller $u$ is due to the high gain.

\begin{figure}[h!]
	\vspace{-0.2cm}	  	
	\includegraphics[width=0.48 \textwidth]{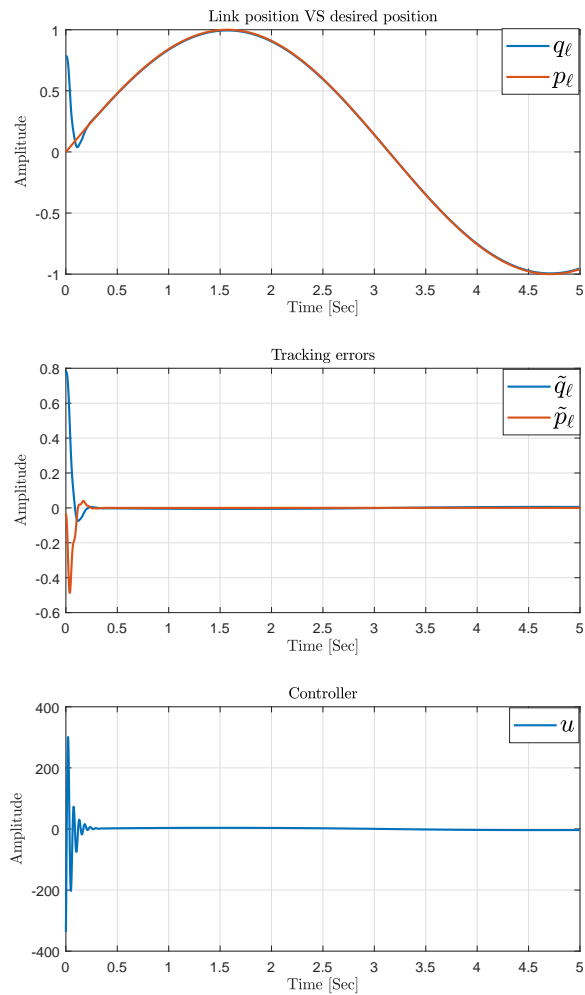}\hfill
	\vspace{-1cm}
	\caption{Performance with joint stiffness $k=3.1$   Nm/rad.}
	\label{fig:positionerror1}	
\end{figure}

If the stiffness constant changes to $k=3.1$, the closed-loop system keeps convergent to the desired steady-state behavior as shown in Figure \ref{fig:positionerror1}. Notice that the transient time is also maintained. However, the control effort has a considerably bigger overshoot than in the previous case.

\section{Conclusion}
In this paper, we propose a virtual differential passivity based control which is applied to the tracking control of FJRs. Firstly, we introduced a virtual system associated to the FJR in the pH framework, which inherits structural properties of the actual system. This  system is used for the control design procedure such that the closed-loop virtual system  is made  strictly differentially passive with a prescribed steady state solution. Furthermore, the closed-loop virtual system preserves the variational  dynamics structure in \eqref{eq:variationalvirtualpHsystem}. We show that the closed-loop virtual system can be seen as the feedback interconnection of two differentially passive subsystems.  The controller $u(x,x,t)$  solves the tracking problem in FJRs. Simulations   confirm the theoretical results. 
A major implementation drawback of our controller presented here is that we require acceleration and jerk measurements. This is  an open problem left for future research. 

\begin{ack}
The first author thanks to Dr. H. Jard\'on-Kojakhmetov for the fruitful discussions that motivated the research on FJRs and to L. Pan for his help in implementing the numerical simulations. The first author is also grateful with CONACyT-Government of the State of Puebla for the scholarship assigned to CVU  $386575$.
\end{ack}

\bibliography{bibliografia-flexible}             
                                                   







\end{document}